\documentclass[12pt]{article}
\begin{document}
\begin{center}
{\Large {\bf Momentum space topological invariants for the $4D$ relativistic vacua with mass gap}}

\vskip-40mm \rightline{\small ITEP-LAT/2012-01 } \vskip 30mm

{%\baselineskip=16pt
\vspace{1cm}
{  M.A.~Zubkov$^{a}$, {G.E. Volovik$^{b,c}$ } }
\\
\vspace{.5cm} {
$^a$ ITEP, B.Cheremushkinskaya 25, Moscow, 117259, Russia\\
$^b$ Low Temperature Laboratory, Aalto University, School of Science and
Technology, P.O. Box 15100, FI-00076 AALTO, Finland
\\
$^c$ Landau Institute for Theoretical Physics RAS, Kosygina 2, 119334 Moscow,
Russia

 } }

\end{center}

\begin{abstract}
Topological invariants for the $4D$ gapped system are discussed with application to the
quantum vacua of relativistic quantum fields. Expression
$\tilde{\cal N}_3$ for the $4D$ systems with mass gap defined in
\cite{Volovik2010} is considered. It is demonstrated that $\tilde{\cal N}_3$
remains the topological invariant when the interacting theory in deep
ultraviolet is effectively massless. We also consider the $5D$ systems and
demonstrate how $4D$ invariants emerge as a result of the dimensional
reduction. In particular, the new $4D$ invariant $\tilde{\cal N}_5$ is
suggested. The index theorem is proved that defines the number of massless
fermions $n_F$ in the intermediate vacuum, which exists at the transition line
between the massive vacua with different values of $\tilde{\cal N}_5$. Namely,
$ 2 n_F$ is equal to the jump $\Delta\tilde{\cal N}_5$ across the transition.
The jump $\Delta\tilde{\cal N}_3$ at the transition determines the number of
only those massless fermions, which live near the hypersurface $\omega=0$.
 The considered invariants are calculated for the lattice model with
Wilson fermions.
\end{abstract}

%\maketitle

\newcommand{\br}{{\bf r}}
\newcommand{\bu}{{\bf \delta}}
\newcommand{\bk}{{\bf k}}
\newcommand{\bq}{{\bf q}}
\def\({\left(}
\def\){\right)}
\def\[{\left[}
\def\]{\right]}

\section{Introduction}

 The momentum space topology is becoming the main tool for the investigation of the robust properties
of ground states (vacua) of condensed matter systems (see review papers
\cite{HasanKane2010,Xiao-LiangQi2011,Volovik2011}). It revealed variety of
vacuum structures with nontrivial topological invariants,
 which is now commonly known as topological materials. These invariants protect gapless fermions in bulk, or on the surface of the fully gapped topological materials, or inside the core of topological defects
 in these systems, such as strings, vortices, monopoles, domain walls, solitons, etc. The gaplessness
(masslessness) of fermionic species is not sensitive to the details of the
microscopic physics:  irrespective of the deformation of the parameters of the
microscopic theory, the energy spectrum of fermions remains strictly gapless.
Though the physics of topological materials started 30 years ago with the
pioneering work by Nielsen and Ninomiya \cite{NielsenNinomiya1981}, who
demonstrated that the masslessness of elementary particles can be protected by
the momentum space topology, the investigation of the topologically nontrivial
vacua in relativistic quantum field theories is still in its infancy, with a
rather few papers  exploiting topological invariants in momentum space (see
e.g.
\cite{So1985,IshikawaMatsuyama1986,Kaplan1992,Golterman1993,Volovik2003,Horava2005,Creutz2008,Kaplan2011}).

Here we are interested in the topologically nontrivial relativistic vacua with
massive fermions. The $4D$ vacua with mass gap may exist in various phases,
which are described by different values of topological invariants.  Recently
such an invariant was considered in \cite{Volovik2010}. This invariant
$\tilde{\cal N}_3$ is determined on the $\omega = 0$ hypersurface of $4D$
momentum space, which represents the $3D$ Brillouin zone. Such invariant is not
sufficient  for the full classification of the topological states of  the
lattice models, whose momentum-frequency $(\omega, {\bf p})$ space represents
the $4D$ Brillouin zone. Here we introduce another topological invariant
$\tilde{\cal N}_5$, which is appropriate for the lattice models (see
Eq.(\ref{N_5})). The invariants $\tilde{\cal N}_3$ and $\tilde{\cal N}_5$ are
then applied to particular lattice model with Wilson fermions (see e.g. Ref.
\cite{Creutz2011}). Table \ref{table2} demonstrates the values of these
invariants in different ranges of mass parameter $m$,  in which the vacuum
represents different phases of topological insulator.

There are crtitical values of mass parameter in the Wilson model $m = 0$, $-2$,
$-4$, $-6$ and $-8$, at which the topological quantum phase transitions occur
between the insulators with different $\tilde{\cal N}_3$ or/and $\tilde{\cal
N}_5$ (more on topological phase transitions, at which the topological charge
of the vacuum changes while the symmetry does not, see \cite{Volovik2007}).  At
these values of $m$ the vacuum states are gapless and they represent analogues
of topological semimetals: similar to the four-dimensional graphene discussed
in Ref. \cite{Creutz2008} they contain massless Dirac fermions.  There is an
analogue of index theorem, which relates the number of these massless fermions
with the jump of the topological invariant across the transition
($\Delta\tilde{\cal N}_5$ or $\Delta\tilde{\cal N}_3$).  The total algebraic
number of gapless fermions at the transition point is  $n_F=\Delta\tilde{\cal
N}_5/2$, and according to Table \ref{table2} the states with critical $m = 0$,
$-2$, $-4$, $-6$ and $-8$  have correspondingly 1, 4, 6, 4, 1 massless species.
The jump $\Delta\tilde{\cal N}_3$ also determines the number of massless
fermions in the intermediate states, but only those of them which live near the
hypersurface $\omega=0$, i.e. $n_F(\omega=0)=\Delta\tilde{\cal N}_3/2$
\cite{Volovik2010}. The intermediate states have correspondingly 1, 3, 3, 1, 0
of such massless species.

Since the invariants are expressed in terms of the Green's function, they are
applicable to the interacting systems as well \cite{Volovik2003}. We prove that
the relation between the number of gapless fermions and the jump in the
topological charge remains valid within the interacting theory for the wide
class of $4D$ models.  (Recent discussion on topological invariants for
interacting condensed matter systems in terms of Green's function and some
peculiarities in the bulk-boundary correspondence there see in Refs.
\cite{Gurarie2011,EssinGurarie2011}.)

%In some special but relevant cases, when, for example, the system is
%effectively massless in ultraviolet (say, for the asymptotic free theory),
%these invariants can be expressed in terms of the Hamiltonian describing the
%asymptotically free fermions in the ultraviolet.

The paper is organized as follows. In Section \ref{SectN3} we consider
topological invariant $\tilde{\cal N}_3$. In Section \ref{Sect5D} we consider
the $5D$ massless systems. In Section \ref{Sect5D4D} we derive topological
invariants in $4D$ via the dimensional reduction of the $5D$ constructions. In
Section \ref{SectN5} we suggest the method of calculation of the invariants
$\tilde{\cal N}_5$ and $\tilde{\cal N}_3$ and formulate the index theorem. In
Section \ref{lattice} we apply the suggested constructions to the lattice model
with Wilson fermions. In Section \ref{SectConcl} we end with the conclusions.

\section{Topological invariant $\tilde{\cal N}_3$}
\label{SectN3}
\subsection{Green functions}

Let us consider for the definiteness $K$ flavors of Dirac fermions $\chi^A, A =
1,..., K$ coupled to some gauge field $\cal A$, that, in turn, may be coupled
to the other unknown fields. Most of the conclusions, however, may be extended
to a more general case. The fermion Green function in Euclidean space has the
form:
\begin{eqnarray}
- i {\cal G}^B_A(x)  &= & \frac{1}{Z} \int D\bar{\chi} D{\cal A}\, {\rm exp}
\Bigl( - S_G[{\cal A}] \nonumber\\ && - \int d^4 x \bar{\chi} [(\partial_i + i
{\cal A}_i) \gamma^i + m] \chi\Bigr) \bar{\chi}_A(0) \chi^B(x)\nonumber\\ &= &
\frac{i}{Z} \int D\bar{\chi} D{\cal A}\, {\rm exp} \Bigl( - S_G[{\cal A}]
\Bigr) \, {\rm Det} \, [i(\partial_i + i {\cal A}_i) \gamma^i + i m] \nonumber\\
&& \{[i(\partial_i + i {\cal A}_i) \gamma^i +  i m]^{-1}\}^B_A \label{Gr}
\end{eqnarray}
Here $S_G$ is some unknown effective action. We use Euclidean $\gamma$ -
matrices such that $\{\gamma^i,\gamma^j\} = 2 \delta^{ij}$. It will also be
implied that a certain Lorentz invariant gauge is fixed. When the interactions
are turned off, we have
\begin{equation}
{\cal G}(\omega, {\bf p})^B_A = \int d^4x {\cal G}^B_A(0,x) e^{i \omega x^4 +i
({\bf p} {\bf x})} = \frac{\delta^B_A}{ p_i \gamma^i - i m }, \, p_4 = \omega
\label{Green0}
\end{equation}

When the interactions are turned on, we suppose that
 Lorentz invariance and  parity are not broken. We shall also use the function ${\cal Q} = -i \gamma^5 {\cal
 G}$ that is Hermitian as follows from (\ref{Gr}). It can be diagonalized with respect to the flavor index via
 Unitary transformations of the fermion field. Later we shall imply that the
 basis is chosen, in which $\cal Q$ (and $\cal G$) is diagonal in flavor index.
 Symmetry fixes the form of the Green function
\begin{equation}
{\cal G}(\omega, {\bf p})^B_A =  {Z^C_A(p^2)}[\frac{1}{ p_i \gamma^i - i m(p^2)
}]^B_C\label{G}
\end{equation}

Here $Z(p^2)$ is the wave function renormalization function, while $m(p^2)$ is
the effective mass term. Both $Z$ and $m$ are diagonal matrices in indices
$A,B$. Below we omit fermion indices $A,B$ in most of the expressions.

In addition we shall also consider more general case, when the Green function
has the form
\begin{equation}
{\cal G}^B_A = \frac{Z[p^2]}{g^i[p] \gamma^i - im[p]}\delta^B_A, \quad i =
1,2,3,4 \label{G_}
\end{equation}
Here $g^i[p]$ is a certain function of the 4 - momentum $p$ that is again
diagonal in flavor indices. We shall also denote $g_5[p] = m[p]$. The Green
function of this form appears, in particular, in the lattice regularization
considered in Section \ref{lattice} when Lorentz symmetry is broken due to the
regularization. It is worth mentioning that (\ref{G_}) is not the general form
of the Green function in the general case when Lorentz symmetry is broken.

%In particular, modern lattice regularizations with lattice chiral symmetry do
%not lead to the Green functions of the form (\ref{G_}) and should be considered
%separately.

As the matrices $g^A_B(p), Z^A_B(p), m^A_B(p)$ are all diagonal we further deal
with them formally as with numbers.
%This form
%of the Green function reflects the possibility of the fermion condensation at
%large $e$.

\subsection{When $\tilde{\cal N}_3$ is the topological invariant?}

In \cite{Volovik2010} the following expression has been considered:
\begin{equation}
\tilde{\cal N}_3 = \frac{1}{24 i \pi^2} {\rm Tr}\, \gamma^5 \gamma^4
\int_{\omega = 0} {\cal G}^{-1} d {\cal G} \wedge d {\cal G}^{-1} \wedge d
{\cal G} \label{N30}
\end{equation}
Here integration is performed along the hypersurface $\omega = 0$ of momentum
space. This is the topological invariant protected by symmetry, i.e. it is
invariant under perturbations only when the Green's function at $\omega=0$ (or
the effective Hamiltonian) commutes with the matrix $\gamma^5 \gamma^4$ (for
topology protected by symmetry see Refs. \cite{Volovik2003,Volovik2011} and
recent paper \cite{Wen2012}). For the noninteracting fermions with positive
masses we have $\tilde{\cal N}_3$  equal to the number of flavors of Dirac
fermions. If the interactions are introduced, the Green function is changed:
${\cal G} \rightarrow {\cal G} + \delta {\cal G}$. If the interactions do not
violate the condition $[{\cal G}(0,{\bf p}), \gamma^5 \gamma^4] = 0$, then
(\ref{N30}) transforms as follows:
\begin{eqnarray}
  \delta \tilde{\cal N}_3  & = &
  \frac{1}{24 i \pi^2} \int_{\omega = 0} {\bf tr} \left(\gamma^5 \gamma^4 \{[\delta {\cal G}]
  d  {\cal G}^{-1}+{\cal G}
  d  [\delta {\cal G}^{-1}]\}\wedge {\cal G}
  d  {\cal G}^{-1}\wedge {\cal G}
  d  {\cal G}^{-1}\right)\nonumber\\&=&
\frac{1}{24 i \pi^2} \int_{\omega=0} {\bf tr} \left(\gamma^5 \gamma^4 \{-{\cal
G} [\delta {\cal G}^{-1}]{\cal G}
  d  {\cal G}^{-1}+{\cal G}
  d  [\delta {\cal G}^{-1}]\}\wedge {\cal G}
  d  {\cal G}^{-1}\wedge {\cal G}
  d  {\cal G}^{-1}\right) \nonumber\\&=&
-\frac{1}{24 i \pi^2} \int_{\omega = 0} {\bf tr} \left(\gamma^5 \gamma^4
\{[\delta {\cal G}^{-1}] [d{\cal G}] +
  d  [\delta {\cal G}^{-1}]{\cal G}\}\wedge
  d  {\cal G}^{-1}\wedge
  d  {\cal G}\right) \nonumber\\&=&
-\frac{1}{24 \pi^2} \int_{\omega=0} d \, {\bf tr} \left(\gamma^5 \gamma^4
\{[\delta {\cal G}^{-1}] {\cal G}] \}
  d  {\cal G}^{-1}\wedge
  d  {\cal G}\right)\nonumber\\&=&
-\frac{1}{24 i \pi^2} \int_{\partial \Sigma}  \, {\bf tr} \left(\gamma^5
\gamma^4 \{[\delta {\cal G}^{-1}] {\cal G}] \}
  d  {\cal G}^{-1}\wedge
  d  {\cal G}\right)\label{dNP}
\end{eqnarray}
Here $\partial \Sigma$ is the boundary of the hypersurface $\omega = 0$ in
momentum space. That's why $\tilde{\cal N}_3$ is the topological invariant if
\begin{equation}
\int_{\partial \Sigma}  \, {\bf tr} \left(\gamma^5 \gamma^4 \, [\delta {\cal
G}^{-1}] {\cal G} \,
  d  {\cal G}^{-1}\wedge
  d  {\cal G}\right)=0 \label{dNP_}
\end{equation}
When the Green function is of the form (\ref{G}), we consider the surface
$\partial \Sigma$ such that ${\bf p}^2 = const \rightarrow \infty$ and denote
$n^i = p^i/|{\bf p}|$). Then Eq. (\ref{dNP_}) is equivalent to
\begin{equation}
\frac{\delta m}{|{\bf p}|\, (1 + \frac{m^2}{p^2})^2}\rightarrow 0, \quad |{\bf
p}| \rightarrow \infty \label{dN3}
\end{equation}
 Thus, we came to the following
\begin{center} {\it Lemma } \end{center}
{\it In the $4D$ system expression (\ref{N30}) is the topological invariant if
$[{\cal G}(0,{\bf p}), \gamma^5 \gamma^4] = 0$ and either momentum space is
closed or on its boundary Eq. (\ref{dNP_}) holds. If the Green function is of
the form of Eq. (\ref{G}) then the latter condition is reduced to Eq.
(\ref{dN3}).}

\subsection{The asymptotic free models}

It is natural to suppose, that in the realistic theory in deep ultraviolet at
$p^2 \rightarrow \infty$ the system becomes effectively massless and chiral
symmetric. We imply that the model is assymptotic free and the fermions are
coupled to the nonabelian gauge field. For a recent numerical investigation of
the fermion propagator in such models see, for example, \cite{Shrock}.
Numerical investigations confirm that $\frac{m(p^2)}{p} \rightarrow 0$ while
$Z(p^2) \rightarrow Z_{\infty}$ at ${\bf p}^2 \rightarrow \infty$. Therefore,
(\ref{dN3}) is fulfiled. Then $\delta \tilde{\cal N}_3 = 0$ and $\tilde{\cal
N}_3 = {\rm Sp} \, {\bf 1} = K$ (the trace here is over the flavor indices) is
equal to the number flavors of Dirac fermions until the phase transition is
encountered. It is worth mentioning that in the case when $Z(p^2) \rightarrow
Z_{\infty}$ and $m(p^2) \rightarrow {\rm const} \, p^2$ at $p^2 \rightarrow
\infty$ we also have $\delta \tilde{\cal N}_3 = 0$ (see \cite{Volovik2010}).

\section{Topological invariants for $5D$ massless fermions}
\label{Sect5D}

\subsection{Topological invariant in $5D$ at $\omega = 0$}

We consider the system with massless fermions. Let us introduce the following
expression:
\begin{equation}
{\cal N}_3^{\prime} = \frac{1}{24 \pi^2 } {\rm Tr}\, \gamma^4 \int_{\omega = 0}
{\cal G}^{-1} d {\cal G} \wedge d {\cal G}^{-1} \wedge d {\cal G} \label{N40_}
\end{equation}
Here integral is over the $3D$ hypersurface that encloses the Fermi point in
momentum space in the $\omega = 0$ subspace. Expression (\ref{N40_}) is
invariant under small perturbations of the Green function if $\{\delta {\cal
G}(0,p), \gamma^4\} = 0$. That's why, the Fermi point in $5D$ is protected by
the given topological invariant if the interactions are such that $\{\delta
{\cal G}(0,p), \gamma^4\} = 0$. For example, if the Green function is of the
form:
\begin{equation}
{\cal G}(\omega, {\bf p}) =  \frac{Z(p^2)}{\omega\gamma^4 +  p_a \gamma^a},
\quad a = 1,2,3,5
\end{equation}
then we obtain  ${\cal N}^{\prime}_3 = 2 \, {\rm Sp}\, {\bf 1}$ (the number of
$5D$ Weil fermions).

\subsection{Topological invariant in $5D$ space $(\omega,{\bf p})$}

Below we generalize the construction of the topological invariant ${\cal
N}^{\prime}_3$ considered above. The resulting construction uses the Green
function defined on the surface that encloses the Fermi point in $\omega - {\bf
p}$ space. Suppose that the $5D$ Green function  has the form
\begin{equation}
 {\cal G}(\omega, {\bf p}) =
\frac{Z(p^2)}{g_a[p] \gamma^a }, \quad a = 1,2,3,4,5, \quad p_4 = \omega
\end{equation}
with real functions $g^a[p]$.

Let us define the function in momentum space
\begin{equation}
{\cal H} = \frac{{\cal G}}{\sqrt{\frac{1}{4}{\rm Tr}\, {\cal G}^+{\cal G}}}
\end{equation}

Here the trace is over the Dirac indices only. We can express ${\cal H}$ as $
{\cal H} = \hat{g}_a \gamma^a,\quad \hat{g}_a = \frac{g_a}{|g|}, \quad |g| =
\sqrt{g_a g_a}, \quad a = 1,2,3,4,5$. Now let us consider the following
integral over the closed $4D$ surface $\Sigma$ in momentum space such that
$\cal G$ does not have poles on $\Sigma$:

\begin{eqnarray}
{\cal N}_4 &=& \frac{3}{16 \pi^2 4!} {\rm Tr}\, \int_{\Sigma} {\cal H}\, d
{\cal H} \wedge d {\cal H}\wedge d {\cal H}\wedge d {\cal H}\nonumber\\& =&
\frac{3}{4 \pi^2 4!} \epsilon_{abcde}\, \int_{\Sigma} \hat{g}^a\, d \hat{g}^b
\wedge d \hat{g}^c \wedge d \hat{g}^d \wedge d \hat{g}^e \label{N4__}
\end{eqnarray}

The given expression for the invariant is reduced to (\ref{N40_}) for the Green
function of the form  $\frac{Z[p^2]}{p^a \gamma^a}$. Without interactions $\cal
G$ has the pole at ${\bf p} = \omega = 0$ that corresponds to the Fermi point.
In this case ${\cal N}_4 = 2 \, {\rm Sp}\, {\bf 1}$ for any surface that
encloses the pole.

\section{Dimensional reduction of the  $5D$ constructions}
\label{Sect5D4D}

\subsection{Dimensional reduction of ${\cal N}_3^{\prime}$}

Let us consider the $4D$ system with the Green function of the form (\ref{G_}).
We have
\begin{equation}
{\cal Q} = \frac{Z[p^2]}{ig^i[p] \gamma^i \gamma^5 + m[p]\gamma^5} = -i
\gamma^5 {\cal G}
\end{equation}
Next, we define new $\gamma$ - matrices $\Gamma^5 = \gamma^5, \, \Gamma^i =
i\gamma^i \gamma^5$ that satisfy $\{\Gamma^a,\Gamma^b\} = 2 \delta^{ab},\, a =
1,2,3,4,5$. We also denote $g_5[p] = m[p]$. Then $\cal Q$ has the form
\begin{equation}
{\cal Q} = \frac{Z[p^2]}{g^a[p] \Gamma^a}, \quad a = 1,2,3,4,5
\end{equation}

Now we can treat $\cal Q$ as a $5D$ Green function defined on the hypersurface
$p_5 = g_5[p_1,p_2,p_3,p_4]$. We may consider, for example, the following
analogue of expression (\ref{N40_}):
\begin{eqnarray}
\tilde{\cal N}_3^{\prime} &=& \frac{1}{24 \pi^2} {\rm Tr}\, \Gamma^4
\int_{\omega = 0} {\cal Q}^{-1} d {\cal Q} \wedge d {\cal Q}^{-1} \wedge d
{\cal Q} \nonumber\\
&=& \frac{1}{24 i \pi^2} {\rm Tr}\, \gamma^5 \gamma^4 \int_{\omega = 0} {\cal
G}^{-1} d {\cal G} \wedge d {\cal G}^{-1} \wedge d {\cal G} \equiv \tilde{\cal
N}_3\label{N411}
\end{eqnarray}

Here integral is over the closed $3D$ hypersurface in the slice $\omega = 0$ of
$4D$ momentum space. One can see that we come to the invariant $\tilde{N}_3$
discussed in Section \ref{SectN3}. This equivalence gives the way to calculate
$\tilde{\cal N}_3$. For example, let us consider the case of noninteracting
massive fermions. We consider the closed surface $\Sigma$ composed of the two
pieces $\Sigma_1: p_5 = m$ and $\Sigma_2: p_5 = - m $ connected at
 infinity by the additional piece of surface. This surface encloses the pole of $\cal Q$ in $5D$ momentum space.
   Then we calculate expression (\ref{N40_}) for this
 surface ${\cal N}^{\prime}_3 = 2 \, {\rm Sp}\, {\bf 1}$.
 Therefore, $\tilde{\cal N}_3 (m) - \tilde{\cal N}_3 (-m)  = 2 \, {\rm Sp}\, {\bf
 1}$ while $\tilde{\cal N}_3 (m) = - \tilde{\cal N}_3 (-m)$. As a result
  $\tilde{\cal N}_3 (m) = {\rm Sp} \, {\rm sign}\, m $.

\subsection{Dimensional reduction of ${\cal N}_4$}

Let us define the function in momentum space
\begin{equation}
{\cal H}^{\prime} = \frac{{\cal Q}}{\sqrt{\frac{1}{4}{\rm Tr}\, ({\cal Q})^2}}
= \hat{g}_a \Gamma^a,\quad \hat{g} = \frac{g}{\sqrt{g^a g^a}}
\end{equation}
(Remind that here $\hat{g}^a$ is the matrix diagonal in flavor indices.)

Now let us consider the following integral over  $4D$ momentum space:
\begin{eqnarray}
{\cal N}^{\prime}_4 &=& \frac{3}{16 \pi^2 4!} {\rm Tr}\, \int {\cal
H}^{\prime}\, d {\cal H}^{\prime} \wedge d {\cal H}^{\prime}\wedge d {\cal
H}^{\prime}\wedge d {\cal H}^{\prime}\nonumber\\& =& \frac{3}{4 \pi^2 4!}
\epsilon_{abcde}\, \int \hat{g}^a\, d \hat{g}^b \wedge d \hat{g}^c \wedge d
\hat{g}^d \wedge d \hat{g}^e \label{N11}
\end{eqnarray}

If momentum space $\cal M$ is closed (for example, has the form of torus or
$4D$ - sphere), then expression (\ref{N11}) is the topological invariant that
measures the degree of mapping $\hat{g} : {\cal M} \rightarrow S^{4}\otimes S^4
... \otimes S^4$. For the calculation of ${\cal N}^{\prime}_4$ in open momentum
space see discussion below in Section \ref{N5calc}.

\subsection{Dimensional reduction of ${\cal N}_5$}

Now let us consider the other way to derive the topological invariant for the
$4D$ gapped systems starting from a $5D$ construction. Let us consider the
Euclidean Green's function on the 4D lattice ${\cal G}$ as the inverse
Hamiltonian in 4D momentum space and introduce the 5D Green's function:
\begin{equation}
\label{Green5}
 G^{-1}(p_5,p_4,{\bf p})= p_5 \gamma^5 + {\cal G}^{-1}(p_4,{\bf p}) = (i p_5  + {\cal Q}^{-1}(p_4,{\bf
 p}))(-i \gamma^5)
\,.
\end{equation}
Then one can introduce the topological invariant as  the 5-form:
\begin{equation}
\label{N_5} {\cal N}_5 = \frac{1}{2 \pi^3 5! i} {\rm Tr}\,  \int G d G^{-1}
\wedge G d G^{-1}\wedge G d G^{-1} \wedge G d G^{-1}\wedge G d G^{-1} \,,
\end{equation}
where the integration is over the  Brillouin zone in 4D momentum space
$(p_4,{\bf p})$ and over the whole $p_5$ axis. In this form the invariant is
applicable also to the interacting case.  Similar to Section \ref{SectN3} we
obtain that  expression (\ref{N_5}) is the topological invariant if the
variation of the Green function $\delta {\cal G}$ satisfies
\begin{equation}
\int_{\partial [{\cal M}\otimes R]}  \, {\bf tr} \left( [\delta {\cal G}^{-1}]
{ G} \,
  d  { G}^{-1}\wedge
  d  { G}\wedge d  { G}^{-1}\wedge
  d  { G}\right)=0,\quad p_5^2 \rightarrow \infty \label{dN5}
\end{equation}
Remarkably, here $\cal G$ may be almost arbitrary. In particular, if $\cal M$
is compact closed space, then (\ref{dN5}) is satisfied for any $\cal G$ that
does not have poles. If $\cal M$ is noncompact, then (\ref{dN5}) is satisfied
for the Green function of the form (\ref{G_}) when the system is massless in
ultraviolet.

In terms of the $4D$ Green function expression (\ref{N_5}) is
\begin{eqnarray}
&& \tilde{\cal N}_5  =  \frac{1}{2 \pi^3 4! i} \int^{\infty}_{-\infty}  d p_5
{\rm Tr}\,  \int_{\cal M} \frac{ 1}{p_5 \gamma^5 + {\cal G}^{-1}} \gamma^5
 \frac{1}{p_5 \gamma^5 + {\cal G}^{-1}}  d {\cal G}^{-1} \nonumber\\ &&
  \wedge \frac{1}{p_5 \gamma^5 + {\cal G}^{-1}} d {\cal G}^{-1}
  \wedge \frac{1}{p_5 \gamma^5 + {\cal G}^{-1}} d {\cal G}^{-1}  \wedge \frac{1}{p_5\gamma^5 + {\cal G}^{-1}}
   d {\cal G}^{-1}  \, \label{N500}
\end{eqnarray}

 In the case, when the 4D Green's
function has the form of Eq. (\ref{G_}): ${\cal G} = \frac{Z[p]}{g_a[p]
\gamma^a -i g_5[p] }$, we denote ${\cal Q} = \frac{1}{f_b[p] \Gamma^b }, b =
1,2,3,4,5$, and $f^a = \frac{g^a}{Z}$. Then
\begin{eqnarray}
 \tilde{\cal N}_5 & = & \frac{1}{2 \pi^3 4! i} {\rm Tr}\,  \int_{\cal M} \int \frac{i d p_5}{(p^2_5 + f^2)^5} (f \Gamma - i p_5)
 \wedge ( - df \Gamma (f^2 + p_5^2) + (f\Gamma - i p_5) d f^2 )  \nonumber\\ && \wedge df \Gamma \wedge
  (f^2 + p_5^2) + (f\Gamma - i p_5) d f^2 ) \wedge df \Gamma \label{N555}
\end{eqnarray}

The integral over $p_5$ is equal to the residue of the corresponding pole at
$p_5 = i f$ that can easily be calculated. As a result invariant (\ref{N555})
after the integration over $p_5$ is reduced to the degree of mapping of the 4D
Brillouin zone to the 4D sphere
 of unit vector:
\begin{equation}
\label{N_51} \tilde{\cal N}_5 = \frac{3}{4 \pi^2 4!} \epsilon_{abcde}\, \int
\hat{g}^a\, d \hat{g}^b \wedge d \hat{g}^c \wedge d \hat{g}^d \wedge d
\hat{g}^e \equiv {\cal N}^{\prime}_4
\end{equation}

We proved here the following
\begin{center} {\it Theorem} \end{center}
{\it Eq. (\ref{N500}) defines the topological invariant for the gapped $4D$
system with momentum space $\cal M$ if Eq. (\ref{dN5}) holds. This requirement
is satisfied, in particular, for the system with compact closed $\cal M$ and
for the system with the Green function of the form (\ref{G}) that is massless
in ultraviolet.

For the system with the Green function of the form (\ref{G_}) expression for
the topological invariant Eg. (\ref{N500}) is reduced to Eq. (\ref{N_51}). }

 The 5-form topological invariant  (\ref{N_5}) has been discussed in \cite{Volovik2003,ZhongWang2010,SilaevVolovik2010}.
 In particular it is responsible for the topological stability of the 3+1 chiral fermions emerging  in the core
  of the domain wall separating  topologically different  vacua in 4+1 systems (see Sec. 22.2.4 in \cite{Volovik2003})
   and fermion zero modes on vortices and strings \cite{SilaevVolovik2010}. The topological invariant
   for the general $2n+1$ insulating relativistic vacua and the bound chiral fermion zero modes emerging
   there have been considered in \cite{Kaplan1992,Golterman1993,Kaplan2011}.

\subsection{Topological invariant $\tilde{\cal N}_4$}

In addition to the invariant $\tilde{\cal N}_5$ let us also consider a
different construction that coincides with  $\tilde{\cal N}_5$ for the case of
free fermions
\begin{equation}
\tilde{\cal N}_4 = \frac{1}{48 \pi^2} {\rm Tr}\, \gamma^5 \int_{\cal M} d{\cal
G}^{-1}\wedge d {\cal G} \wedge d {\cal G}^{-1} \wedge d {\cal G} \label{N40}
\end{equation}
Here the integration is over the whole $4D$ space $\cal M$. The expression in
this integral is the full derivative. Therefore, we have:
\begin{eqnarray}
\tilde{\cal N}_4 &=& \frac{1}{48 \pi^2} {\rm Tr}\, \gamma^5 \int_{\cal M}
d{\cal G}^{-1}\wedge
d {\cal G} \wedge d {\cal G}^{-1} \wedge d {\cal G} \nonumber \\
& = & \frac{1}{48 \pi^2} {\rm Tr}\, \gamma^5 \int_{\partial {\cal M}} {\cal
G}^{-1} d {\cal G} \wedge d {\cal G}^{-1} \wedge d {\cal G}
\end{eqnarray}
Here the integral is over the $3D$ hypersurface $\partial {\cal M}$. The last
equation is identical (up to the factor $1/2$) to that of for the invariant
${\cal N}_3$ for massless fermions. Therefore, Eq. (\ref{N40}) defines the
topological invariant if the Green function anticommutes with $\gamma^5$ on the
boundary of momentum space.

For the noninteracting fermions
\begin{equation}
{\cal G}(\omega, {\bf p}) \rightarrow \frac{1}{ p_i \gamma^i }, \, p^2
\rightarrow \infty
\end{equation}
Therefore, $\tilde{\cal N}_4 = {\rm Sp}\, {\bf 1}$ (the number of Dirac
fermions). The value of the given invariant coincides with $\tilde{\cal N}_5$
in this case. However, it will be shown in Section \ref{lattice} that in the
other nontrivial cases like the lattice regularized models the two invariants
give different values. In particular, $\tilde{\cal N}_4 = 0$  while
$\tilde{\cal N}_5$ may be different from zero for the systems with closed
momentum space $\cal M$.

When the interactions are turned on, but in the deep ultraviolet at $p^2
\rightarrow \infty$ the system becomes effectively massless and chiral
symmetric, then  ${\cal G}(p)$ anticommutes with $\gamma^5$ at $p^2 \rightarrow
\infty$. As a result, the value of $\tilde{\cal N}_4$ remains equal to the
number of Dirac fermions until a phase transition is encountered.

\section{Topological invariant $\tilde{\cal N}_5$}
\label{SectN5}

\subsection{A way to calculate  $ \tilde{\cal N}_5$}
\label{N5calc}

In this subsection we consider the Green function of the form (\ref{G_}).
 Let us introduce the following parametrization
\begin{equation}
\hat{g}_5 = {\rm cos} 2 \alpha, \quad \hat{g}_a = k_a {\rm sin} 2 \alpha
\end{equation}

$k$ may be undefined at the points of momentum space ${\cal M}$, where
$\hat{g}^a = 0, a = 0,2,3,4$. In nondegenerate case this occurs on points $p_i,
i = 1, ...$. We denote the small vicinity of the regions where $k$ is undefined
by $\Omega = \Omega(p_0) + \Omega(p_1) + ...$. Then the expression for
$\tilde{\cal N}_5$ can be rewritten as follows:
\begin{eqnarray}
\tilde{\cal N}_5 &=& \frac{3}{ \pi^2 4!} \epsilon_{abcd}\, \int_{{\cal
M}-\Omega} {\rm sin}^3 2 \alpha d
\alpha \wedge k^a\, d k^b \wedge d k^c \wedge d k^d   \nonumber\\
&=& -\frac{1}{ \pi^2 4!} \epsilon_{abcd}\, \int_{\partial {\cal M} -
\partial \Omega} (3 {\rm cos} 2 \alpha - {\rm cos}^3 2 \alpha)  k^a\, d k^b \wedge
d k^c \wedge d k^d \nonumber\\
&=& -\frac{1}{ \pi^2 4!} \epsilon_{abcd}\, \int_{\partial {\cal M} -
\partial \Omega} (3 \hat{g}_5 - \hat{g}_5^3 )  k^a\, d k^b \wedge
d k^c \wedge d k^d, \, \quad k^a = \frac{g^a}{\sqrt{g^a g^a}}
\end{eqnarray}

For the convenience we denote all boundary of momentum space (that may be
consisted of several pieces) by $\partial {\cal M} = -
\partial \Omega(p_{\infty})$. This allows to rewrite the last expression as:
\begin{eqnarray}
\tilde{\cal N}_5 &=&  \frac{1}{ \pi^2 4!} \sum_{i = 0,1,...,\infty}
\epsilon_{abcd}\, \int_{\partial \Omega(p_i)} (3 \hat{g}_5 - \hat{g}_5^3 )
k^a\, d k^b \wedge d k^c \wedge d k^d \label{N5000}
\end{eqnarray}

Let us consider the case when $\hat{g}_5 = 0$ on $\partial {\cal M}$. In
particular, if the boundary of momentum space is placed at infinity and the
system is effectively massless in ultraviolet,  then $\hat{g}^5[p] = 0$ at
$|p|\rightarrow \infty$. We also take into account that $\hat{g}_5 = \pm 1$ at
$p_i, i \ne \infty$.  That's why we obtain:
\begin{eqnarray}
\tilde{\cal N}_5 &=&  \frac{2}{ \pi^2 4!} \sum_{i = 0,1,...} \epsilon_{abcd}\,
\int_{\partial \Omega(p_i)} {\rm sign} \,  (\hat{g}_5) \, k^a\, d k^b \wedge d
k^c \wedge d k^d \label{N5f}
\end{eqnarray}
Here the sum over $p_i$ does not include $p_{\infty}$.

In the case when there is only one zero of $\hat{g}_a$ (we call this point
$p_0$) we also obtain:
\begin{eqnarray}
\tilde{\cal N}_5 &=&  \frac{2}{ \pi^2 4!} \epsilon_{abcd}\, \int_{\partial
\Omega(p_0)} {\rm sign} \,  (\hat{g}_5) \,
k^a\, d k^b \wedge d k^c \wedge d k^d \nonumber\\
&=&  \frac{2}{ \pi^2 4!} \epsilon_{abcd}\, \int_{\partial {\cal M}} {\rm sign}
\,  (\hat{g}_5) \, k^a\, d k^b \wedge d k^c \wedge d k^d
\end{eqnarray}

Suppose that the effective mass $m[p]$ does not change sign anywhere. Then in
this case $\tilde{\cal N}_5$ is equal to the degree of mapping
\begin{equation}
k^a = \frac{g^a(p)}{\sqrt{[g^1(p)]^2 + [g^2(p)]^2 + [g^3(p)]^2 + [g^4(p)]^2}}:
S^3(|p| \rightarrow \infty) \rightarrow S^3\otimes ... \otimes S^3
\end{equation}

Here we proved the following
\begin{center} {\it Theorem} \end{center}
{\it In the system with the Green function of the form (\ref{G_}) the
topological invariant $\tilde{\cal N}_5$ is given by the sum in Eq. (\ref{N5f})
if $\hat{g}_5[p] = 0$ on the boundary of momentum space.}

When the Lorentz symmetry is not broken at $|p| \rightarrow \infty$, and
$g^a[p] \sim  p^a$, we have $\tilde{\cal N}_5 = {\rm Sp}\, {\rm sign}\, m$,
where $m$ is the mass.  That's why for the free fermions with positive masses
$\tilde{\cal N}_5$ is equal to the number of the flavors of Dirac fermions.

The situation, when several points $p_i$ appear within the Brillouine zone will
be considered later when we shall calculate $\tilde{\cal N}_5$ for the lattice
fermions in Section \ref{lattice}.

It is worth mentioning that the method suggested above could be easily extended
to the calculation of $\tilde{\cal N}_3$.

\subsection{Index theorem}
\label{SectIndTheor}

 In this subsection we prove the following

\begin{center} {\it Theorem} \end{center}
{\it Suppose that the $4D$ system depends on parameter $\beta$ and there is a
phase transition at $\beta_c$ with changing of $\tilde{\cal N}_5$. We denote
$\tilde{\cal N}_5 = n_+$ for $\beta > \beta_c$ and $\tilde{\cal N}_5=n_-$ for
$\beta < \beta_c$. In addition we require that $\cal G$ does not have zeros and
does not contain $\gamma_5$ matrix (i.e. ${\rm Tr}\, {\cal G}^{-1}\gamma^5 =
0$). Momentum space $\cal M$ of the $4D$ model is supposed to be either compact
and closed or open. In the latter case we need that $\cal G$ does not depend on
$\beta$ on $\partial {\cal M}$. Then at $\beta = \beta_c$ there are $n_f = (n_+
- n_-)/2$ flavors of massless Dirac fermions. }

Let us consider $6D$ space $(p_1,p_2,p_3,p_4,p_5,p_6)$ with $p_6 = \beta$. $4D$
subspace $p_1,p_2,p_3,p_4$ has the form of the $4D$ Brillouin zone $\cal M$
while $p_5$ axis and $p_6$ axis are straight lines. We then consider the $5D$
hypersurfaces $\Sigma_+$ at $p_6 = \beta_+ > \beta_c$ and $\Sigma_-$ at $p_6 =
\beta_- < \beta_c$. These surfaces, in turn, contain the Brilloiun zones of the
initial model at $p_5 = 0$ for $\beta = \beta_+, \beta_-$ respectively. We
connect $\Sigma_{\pm}$ at $p_5 = \pm \infty$ and at the boundary of $\cal M$
(this is possible because $\cal G$ does not depend on $\beta $ at $\partial
{\cal M}$). The resulting hypersurface $\Sigma$ is closed in $6D$ space.

Let us construct the $6D$ Green function
\begin{equation}
\label{Green6}
 G^{-1}(p_6,p_5,p_4,{\bf p})= p_5 \gamma^5 + {\cal G}^{-1}(\beta,p_4,{\bf p}),
 \quad p_6 = \beta
\end{equation}

Then we consider the invariant
\begin{equation}
\label{N_56} {\cal N}_5|_{\Sigma} = \frac{1}{2 \pi^3 5! i} {\rm Tr}\,
\int_{\Sigma} G d G^{-1} \wedge G d G^{-1}\wedge G d G^{-1} \wedge G d
G^{-1}\wedge G d G^{-1} \,,
\end{equation}
We may neglect in this integral the region placed at infinite $p_5$. That's why
\begin{equation}
{\cal N}_5|_\Sigma = \tilde{\cal N}_5(\beta_+) - \tilde{\cal N}_5(\beta_-) =
n_+ - n_-
\end{equation}

On the other hand, we may deform $\Sigma$ in such a way that it is not already
placed at infinity. ${\cal N}_5|_\Sigma = n_+ - n_-$ means that the $6D$ system
described by  Green function (\ref{Green6}) has $(n_+ - n_-)/2$ massless Dirac
fermions.

$\cal G$ does not contain $\gamma_5$ matrix. Therefore, there are no poles of
$G$ at nonzero $p_5$. That's why we may deform $\Sigma$ in such a way that on
this hypersurface $p_5 = \pm \epsilon$ with arbitrary small $\epsilon$. Also we
make $\beta_{\pm}$ infinitely close to $\beta_c$. As a result the $6D$ model
with Green function (\ref{Green6}) and $(n_+ - n_-)/2$ flavors of massless
Dirac fermions is dimensionally reduced to the $4D$ system with $\beta =
\beta_c$. There are $(n_+ - n_-)/2$ poles in (\ref{Green6}) at $p_6 = \beta_c,
p_5 =0$. We conclude, therefore, that there are $(n_+ - n_-)/2$ poles in ${\cal
G}(\beta_c, p_4, {\bf p})$.

The example of the $4D$ system with compact momentum space that satisfies the
given theorem is given by the model with Wilson fermions considered in Section
\ref{lattice}. The example of the system with noncompact momentum space is
given by the free continuum fermions (in this case mass plays the role of
parameter $\beta$).

It is worth mentioning that a similar theorem can be proved that relates the
jump in $\tilde{\cal N}_3$ with the number of massless fermions that exist in
the $\omega = 0$ hypersurface of the Brillouin zone within the system at the
interface between two gapped states. (Illustration of this theorem see also in
 Section \ref{lattice}).

\section{Momentum space topology of lattice regularized model}
\label{lattice}

\subsection{$\tilde{\cal N}_3$ for free Wilson fermions}

In this section we consider the case of the single flavor of lattice Dirac
fermions. In lattice regularization the periodic boundary conditions are used
in space direction and antiperiodic boundary conditions are used in the
imaginary time direction. The momenta to be considered, therefore, also belong
to a lattice:
\begin{equation}
p_a = \frac{2\pi K_a}{N_a}\,  \quad p_4 = \frac{2\pi K_4+\pi}{N_t}, \quad
K_a,K_4 \in Z\quad a = 1,2,3
\end{equation}
Here $N_a, N_t$ are the lattice sizes in $x$, $y$, $z$, and imaginary time
directions, correspondingly. However, for the lattice of infinite volume we
recover continuous values of momenta that belong to the $4D$ torus.

Let us consider the simplest regularization of fermion action called Wilson
fermions. In the absence of interactions the Green function has the form:

\begin{eqnarray}
{\cal G}& = & \Bigl( \sum_a \Gamma_a {\rm sin}\, p_a - i (m + \sum_a (1 - {\rm
cos}\, p_a)) \Bigr)^{-1}\nonumber\\ & = & \frac{ \sum_a \Gamma_a {\rm sin}\,
p_a + i (m + \sum_a (1 - {\rm cos}\, p_a)) }{\sum_a {\rm sin}^2\, p_a + (m +
\sum_a (1 - {\rm cos}\, p_a))^2}, \quad a = 1,2,3,4
\end{eqnarray}

In Table \ref{table} we represent the spectrum of the model for different
values of $m$.
\begin{table}
\begin{center}
\begin{small}
\begin{tabular}{|c|c|c|c|c|c|c|c|c|c|c|c|c|c|c|c|c|}
\hline
$m = M a $ & $M^1$ & $(2/a + M)^4$ & $(4/a + M)^6$ &$(6/a + M)^4$ &$(8/a + M)^1$  \\
\hline
$m = -2 + M a$ & $(-2/a + M)^1$ & $M^4$ & $(2/a + M)^6$ &$(4/a + M)^4$ &$(6/a + M)^1$  \\
\hline
$m = -4 + M a$ & $(-4/a + M)^1$ & $(-2/a +M)^4$ & $M^6$ &$(2/a + M)^4$ &$(4/a + M)^1$  \\
\hline
$m = -6 + M a$ & $(-6/a + M)^1$ & $(-4/a +M)^4$ & $(-2/a +M)^6$ &$M^4$ &$(2/a + M)^1$  \\
\hline
$m = -8 + M a$ & $(-8/a + M)^1$ & $(-6/a +M)^4$ & $(-4/a +M)^6$ &$(-2/a +M)^4$ &$ M^1$  \\
\hline
\end{tabular}
\end{small}
\end{center} \caption{The spectrum of the system with free Wilson fermions.
In the first column the values of $m$ are specified. In
the other columns masses of the doublers are listed. Everywhere it is implied
that $|Ma| << 1$, where $a$ is the lattice spacing. Expression ${v}^x$ means
$x$ states with masses equal to $v$. } \label{table}
\end{table}
The diagonal values in this table represent physical massive states while the
off - diagonal elements of the table represent unphysical doublers with masses
that tend to infinity in the limit $a \rightarrow 0$. Due to the periodical
boundary conditions $\tilde{\cal N}_3$ remains the topological invariant on the
torus. In Table \ref{table2} we represent the values of $\tilde{\cal N}_3$
versus the values of $m$. Here ${\tilde{\cal N}}_3 = \sum_{i=0}^{3} (-1)^i
C_3^i \frac{m+2i}{|m+2i|}$, where the sum is over the fermion doublers in $3D$,
$m + 2i$ is the mass of the $i$-th doubler while $C_3^i$ is its degeneracy (for
the derivation see \cite{Golterman1993}). It is worth mentioning that on the
given lattice $\tilde{\cal N}_4 = 0$. The value of $\tilde{\cal N}_4$ may be
different from zero if $\cal G$ has poles or zeros somewhere in $4D$ momentum
space.

\begin{table}
\begin{center}
\begin{small}
\begin{tabular}{|c|c|c|c|c|c|c|c|c|c|c|c|c|c|c|c|c|}
\hline
$m$ & $\tilde{\cal N}_3$ & $\tilde{\cal N}_5$   \\
\hline
$m>0 $ & $0$ & $0$ \\
\hline
$-2 < m < 0$ & $-2$ & $-2$  \\
\hline
$-4 < m < -2 $ & $4$ & $6$ \\
\hline
$-6 < m < -4$ & $-2$ & $-6$  \\
\hline
$-8 < m < -6 $ & $0$ & $2$  \\
\hline
$ m < -8 $ & $0$  & $0$ \\
\hline
\end{tabular}
\end{small}
\end{center}
\caption{The values of topological invariants $\tilde{\cal N}_3$ and
$\tilde{\cal N}_5$ for free Wilson fermions. } \label{table2}
\end{table}

Let us notice that the values of the considered topological invariants on the
lattice with periodical boundary conditions contradict to the continuum result.
This is due to the fermion doublers that give essential contributions in spite
of their unphysically large masses.

\subsection{$\tilde{\cal N}_5$ for free Wilson fermions}

As follows from Table \ref{table2}, the invariant $\tilde{\cal N}_3$ does not
resolve between the fully gapped states in the region $-8<m<-6$ and in the
region $m<-8$, which are separated by the gapless state at $m=-8$ and thus
should be topologically different. The reason for that is that the invariant
$\tilde{\cal N}_3$ is determined only at $\omega=0$, while at $m=-8$ the only
gapless state has nonzero $\omega$ (that is $\cos p_4=-1$). Since at $m=-8$
there are no gapless fermions with $\omega=0$, there is no jump in $\tilde{\cal
N}_3$ across this transition. However, the  topological invariant, which is
determined for all $\omega$, may differentiate between the two states with
$\tilde{\cal N}_3=0$. The invariant $\tilde{\cal N}_5$ plays such a role, and
one has the index theorem (see Section \ref{SectIndTheor}), which states that
the total number $n_F$ of gapless fermions emerging at the critical values of
mass  $m$ is determined by the jump in $\tilde{\cal N}_5$. While the jump
$\Delta\tilde{\cal N}_3$ across the transition determines  the number of those
gapless fermions which live in the vicinity of zero frequency, $n_F(\omega=0)$.
The resulting index theorems read:
\begin{equation}
\label{IndexTheorem}
n_F=\frac{1}{2}\Delta\tilde{\cal N}_5~~,~~n_F(\omega=0)=\frac{1}{2}\Delta\tilde{\cal N}_3 \,.
\end{equation}
According to Section \ref{SectIndTheor} this equation is valid for the
interacting systems too. In some cases, however, when momentum space has a
complicated form, one must take into account not only poles of the Green's
function of massless fermions but also zeroes
\cite{Gurarie2011,SilaevVolovik2012}.

For the lattice model with Wilson fermions we use the method described in
Section \ref{N5calc} in order to calculate the invariant $\tilde{\cal N}_5$.
One obtains that the zeros of $\hat{g}^a$ appear at the positions of the
fermion doublers. The corresponding values of ${\rm sign} \, \hat{g}_5$
coincide with the signs of the masses of the doublers. The positions of
doublers are $p_{n_i} = (\pi n_1, \pi n_2, \pi n_3, \pi n_4), \quad n_i = 0,1$.
We also have $\partial_b \hat{g}^a(p_{n_i}) \sim (-1)^{n_a}
\delta^a_b\frac{1}{|m_{n_i}|}$, where $m_{n_i}$ is the mass of the doubler.
That's why, using expression (\ref{N5f}) we obtain:
\begin{equation}
\tilde{\cal N}_5={\cal N}^{\prime}_4 = \sum_{k=0}^{4} (-1)^k C_4^k
\frac{m+2k}{|m+2k|},
\end{equation}
where the sum is over the fermion doublers in $4D$, $m + 2k$ is the mass of the
$k$-th doubler while $C_4^k$ is its degeneracy. This is similar to the result
obtained in \cite{Golterman1993} for the different space-time dimension. The
values of the invariant $\tilde{\cal N}_5$ as a function of bare mass $m$ are
also represented in Table \ref{table2}. The index theorem
Eq.(\ref{IndexTheorem}) gives at the transition points
 $m = 0$, $-2$, $-4$, $-6$,$-8$ the number of gapless fermions correspondingly $n_F=1, 4, 6, 4, 1$
 and $n_F(\omega=0)=1, 3, 3, 1, 0$.

\subsection{Turning on interactions}

When the interaction of Wilson fermions with the lattice gauge field ${\cal U}
= e^{i {\cal A}}$ defined on links is turned on, we have (in coordinate space):

\begin{eqnarray}
&&{\cal G}(x,y)  =  \frac{i}{Z} \int  D{\cal U}\, {\rm exp} \Bigl( - S_G[{\cal
U}] \Bigr) \, {\rm Det}  ({\cal D}[{\cal U},m]) {\cal D}_{x,y}^{-1}[{\cal U},m]
\label{GrU}
\end{eqnarray}
where $S_G$ is the gauge field action while
\begin{equation}
{\cal D}_{x,y}[{\cal U},m]  =  - \frac{1}{2}\sum_i [(1 +
\gamma^i)\delta_{x+{\bf e}_i, y} {\cal U}_{x+{\bf e}_i, y}  +  (1 -
\gamma^i)\delta_{x-{\bf e}_i, y} {\cal U}_{x-{\bf e}_i, y}] +  (m + 4)
\delta_{xy}
\end{equation}

Here ${\bf e}_i$ is the unity vector in the $i$ - th direction. We expect that
the Green function in momentum space has the form \cite{Shrock}:
\begin{eqnarray}
{\cal G}& = & \frac{Z[p]}{ \sum_a \Gamma_a g_a[p] - i m[p]  }
\end{eqnarray}
with unknown functions $Z[p]$, $g_a[p]$, and $m[p]$. That's why the Green
function for the interacting Wilson fermions has the form (\ref{G_}) as well as
the Green function for the noninteracting fermions.

Free Wilson fermions exist in several phases for different values of $m$. These
phases are marked by the values of $\tilde{\cal N}_5$. When the interactions
are turned on, the value of $\tilde{\cal N}_5$ remains equal to its initial
number until the phase transition is encountered. This is related to the fact
that zeros of $g_a[p]$ cannot disappear without changing the phase structure of
the model. These zeros, in turn, are related to the value of $\tilde{\cal N}_5$
according to (\ref{N500}). It is worth mentioning that in order to consider
interactions with nonabelian gauge fields we must consider several flavors of
the fermions as it was done in the previous sections. All results obtained in
the present section may be generalized to this case in a straightforward way.

\section{Conclusions and discussion}
\label{SectConcl}

In the present paper we discuss the topological invariants in momentum space of
the $4D$ models with the mass gap. In particular, the expression for
$\tilde{\cal N}_3$ introduced in \cite{Volovik2010} is considered. We find the
conditions under which $\tilde{\cal N}_3$ is the topological invariant. We also
introduce new functional $\tilde{\cal N}_5$ that is represented as an integral
over the whole $4D$ momentum space. $\tilde{\cal N}_5$ is obtained via the
dimensional reduction of the $5D$ construction. In the form of Eq. (\ref{N500})
this functional appears to be the topological invariant for the wide class of
models. In particular, it is the invariant for the asymptotic free gauge
theories and for the models with compact closed momentum space. In addition, we
prove the index theorem that defines the number of massless fermions $n_F$ in
the intermediate vacuum, which exists at the transition line between the
massive vacua with different values of $\tilde{\cal N}_5$.

In order to illustrate the properties of the mentioned constructions we
consider the lattice model with Wilson fermions.  We found that the vacuum
states of lattice models with fully gapped fermions (insulating vacua) in 4D
space-time are characterized by two topological invariants, $\tilde{\cal N}_3$
and $\tilde{\cal N}_5$. Due to the index theorem mentioned above, they are
responsible, in particular, for the number of gapless fermions which appear at
the topological transition between the massive states with different
topological charges. The vacuum in the intermediate state represents the
topological semimetal. According to Eq.(\ref{IndexTheorem}) the jump  in
$\tilde{\cal N}_5$ determines the total number of massless fermions in the
intermediate state, while the jump in $\tilde{\cal N}_3$ gives the number of
those massless fermions, which have $\omega=0$.

 The simple pattern described above may have a certain connection to the unification of
  fundamental interactions. Namely, masses of all elementary
particles are extremely small compared to the Planck scale, which may indicate
that we  live in the vicinity of the critical line of quantum phase transition
like the transition  between the two gapped states in the model with Wilson
fermions. There is a special reason why nature may choose the vicinity of the
transition line: the massless fermions in the intermediate vacuum states are
able to accommodate more entropy than the gapped states \cite{Volovik2010}.

It would be interesting to apply the analysis similar to the presented here to
the consideration of the other lattice models with more complicated structure
of momentum space (for example, to the overlap fermions \cite{Overlap}).

Also it would be interesting to consider lattice models that describe vacua
with broken symmetry, such as superconducting vacuum in Ref.
\cite{Chernodub2012}, technicolor models with spontaneously broken chiral
symmetry etc. However, such an analysis may necessarily involve higher Green
functions because the two - point functions only do not describe appearance of
the condensates completely.

The work of M.A.Z. was partly supported by RFBR grants 09-02-00338,
11-02-01227, by Grant for Leading Scientific Schools 679.2008.2, by the Federal
Special-Purpose Programme 'Cadres' of the Russian Ministry of Science and
Education, by Federal Special-Purpose Programme 07.514.12.4028.
The work of G.E.V. is supported in part by the Academy of Finland and its COE program.

\end{document}